\pdfoutput=1
\documentclass[times, preprint, 10pt]{elsarticle}
\usepackage{microtype}
\usepackage{graphicx}
\usepackage{booktabs}
\usepackage{xurl}
\usepackage{hyperref}
\hypersetup{hidelinks}
\usepackage{amsmath}
\usepackage{amssymb}
\usepackage{mathtools}
\usepackage{amsthm}
\usepackage{xcolor}
\usepackage{multirow}
\usepackage{colortbl}
\usepackage{tabularx}
\usepackage[capitalize,noabbrev]{cleveref}
\usepackage{xspace}
\biboptions{numbers,sort&compress}
\begin{document}
\begin{frontmatter}

\title{Semantic-Aware Deformable Feature Alignment for Thyroid Nodule Segmentation in Ultrasound}

\author[1]{Lei Zhang\fnref{equal}}
\author[2]{Ruhuan Zhou\fnref{equal}}
\author[1]{Busheng Tong}
\author[3]{Dan Lv}
\ead[url]{https://orcid.org/0000-0002-9288-0863}
\author[1]{Bingyu Liang}
\author[1]{Taibao Wang}
\author[1]{Xinyu Shi}
\author[5]{Yifan Gao}
\author[4]{Ming Zhang\corref{cor1}}
\ead{zhangming@cuhk.edu.cn}
\ead[url]{https://orcid.org/0000-0002-5251-5810}
\author[1]{Qin Wang\corref{cor2}}
\ead{wangqin@ahmu.edu.cn}
\ead[url]{https://orcid.org/0000-0002-2095-6216}

\cortext[cor1]{Corresponding author.}
\cortext[cor2]{Corresponding author.}
\fntext[equal]{These authors contributed equally to this work.}

\affiliation[1]{organization={Department of Otolaryngology, Head and Neck Surgery, The First Affiliated Hospital of Anhui Medical University}, addressline={218 Jixi Road}, city={Hefei}, state={Anhui}, postcode={230031}, country={China}}
\affiliation[2]{organization={Department of Otolaryngology, Head and Neck Surgery, The Second People's Hospital of Hefei, Hefei Hospital Affiliated to Anhui Medical University}, addressline={246 Heping Road}, city={Hefei}, state={Anhui}, postcode={230032}, country={China}}
\affiliation[3]{organization={Department of Otorhinolaryngology, Head and Neck Surgery, West China Hospital, Sichuan University}, addressline={37 Guoxue Lane}, city={Chengdu}, state={Sichuan}, postcode={610041}, country={China}}
\affiliation[4]{organization={Anatomy, School of Medicine, The Chinese University of Hong Kong (Shenzhen)}, addressline={2001 Longxiang Road, Longgang District}, city={Shenzhen}, postcode={518100}, country={China}}
\affiliation[5]{organization={School of Biomedical Engineering, University of Science and Technology of China}, city={Hefei}, state={Anhui}, postcode={230026}, country={China}}

\begin{abstract}
Accurate segmentation of thyroid nodules in ultrasound images is essential for clinical diagnosis yet remains challenging due to boundary ambiguity, speckle noise, and complex geometric variability. Existing methods often fail to reconcile global semantic understanding with precise local boundary delineation, leading to errors in irregular cases. To address this, we propose SDFA-Net, a framework built on semantic-aware deformable feature alignment that explicitly decouples the learning of robust global semantics and detailed local geometry. Our approach integrates three main components: a clustered linear context aggregation module for efficient long-range dependency modeling, an affinity-driven token modulation mechanism using dynamic prototypes to suppress background interference, and a cascaded spatially-aligned feature fusion strategy that achieves sub-pixel geometric alignment across multiple resolution stages. By filtering and aligning multi-scale features, SDFA-Net ensures robust fusion that is both semantically meaningful and geometrically accurate. We validate our method through extensive experiments on three public thyroid ultrasound datasets, TN3K, TUCC, and DDTI, where it demonstrates superior performance over state-of-the-art methods. The proposed framework offers a generalizable solution for precise medical image segmentation in boundary-ambiguous scenarios.
\end{abstract}

\begin{keyword}
Thyroid nodule segmentation \sep deformable alignment \sep prototype learning \sep feature fusion \sep ultrasound imaging
\end{keyword}

\end{frontmatter}

\section*{Glossary}

\noindent\begin{tabularx}{\linewidth}{@{}p{0.12\linewidth}X@{}}
\toprule
Term & Definition \\
\midrule
CNN & Convolutional neural network, used for high-resolution local feature extraction. \\
ViT & Vision Transformer, an architecture that models relations between image patches. \\
CLCA & Clustered Linear Context Aggregation, the global-context module in SDFA-Net. \\
DCN & Deformable convolutional network, which learns spatial sampling offsets. \\
BCE & Binary cross-entropy, a pixel-level segmentation loss. \\
DSC & Dice Similarity Coefficient, a region-overlap measure. \\
HD95 & 95th percentile Hausdorff distance, a boundary-distance measure that reduces the influence of extreme outliers. \\
GFLOPs & Billions of floating-point operations, used to report computational cost. \\
FPS & Frames per second, used to report inference speed. \\
\bottomrule
\end{tabularx}

\section{Introduction}

Thyroid nodules have become increasingly prevalent globally, making accurate diagnosis via ultrasound imaging a clinical priority \cite{Ha2019_9fbe}. Accurate nodule segmentation is essential for computer-aided diagnosis, as it enables the quantitative assessment of size, shape, and texture. However, ultrasound characteristics such as low contrast, speckle noise, acoustic shadowing, and foreground-background feature overlap make nodule boundaries ambiguous and difficult to delineate \cite{Bass2021_38ac,Zhang2026SDCNet}. Furthermore, nodules exhibit significant geometric variability, ranging from well-circumscribed oval masses to irregular, spiculated structures \cite{li2024ultrasound,xie2016ultrasonography}. This variability requires a segmentation model to integrate a global understanding of the nodule's semantic context—such as its overall shape and location—with local geometric details to trace its boundary.

Convolutional Neural Networks (CNNs), particularly U-Net and its variants, are the standard for this task \cite{Ronneberger2015_e91a,gao2021gomokunet}. These networks capture local spatial patterns and textures effectively but are often constrained by limited receptive fields, which can hinder global reasoning and the understanding of large-scale anatomical structures \cite{Loos2024_19e1}. While advancements like dilated convolutions and pyramid pooling modules attempt to expand the receptive field, they often struggle to balance local detail and global context without sacrificing resolution or introducing excessive parameters \cite{Chen2018_09f5}.

To address CNN limitations, recent works have used self-attention, recurrent feature interaction, and pooling-based token aggregation to capture long-range dependencies \cite{Dosovitskiy2020_b18a,Yang2026CCPNet,Li2026RecA}. Hybrid architectures such as TransUNet \cite{Chen2021_7e30} and Swin-UNet \cite{Cao2021_9b92} model global anatomical context using self-attention. However, these approaches often mix contextual information from the entire image, making them susceptible to noise from surrounding tissues and artifacts \cite{Chen2024_1ee2}. Moreover, standard fusion strategies typically rely on simple concatenation or element-wise addition, assuming a fixed spatial correspondence between global and local feature maps. This assumption fails when geometric misalignment occurs due to the irregular shapes of thyroid nodules, leading to suboptimal feature integration \cite{Dai2017_a4a3,Jaderberg2015_9523}.

Global context and fine boundary evidence enter segmentation networks at different resolutions. Their fusion is difficult in thyroid ultrasound because speckle and tissue similarity weaken semantic cues, while irregular nodules shift boundary features across scales. Global attention can collect unrelated tissue responses, and fixed grid fusion assumes a spatial match that the data do not provide. The resulting pattern recognition problem is to select task relevant context and register it with local geometry before prediction.

SDFA-Net addresses this problem through a filter and align pipeline. Clustered linear context aggregation captures long range structure at linear complexity. Dynamic foreground and background prototypes score the semantic relevance of global tokens and suppress responses from surrounding tissue. A cascaded deformable fusion module then warps the filtered context toward high resolution boundary features at three scales. The resulting representation carries global nodule identity and local shape evidence in the same spatial frame.

Experiments on TN3K, TUCC, and DDTI test the three claims behind this design. Component studies measure the contribution of context aggregation, prototype filtering, and spatial alignment. Comparisons with ten segmentation models test region accuracy and boundary quality through Dice and HD95. Shape and size analysis examines the setting in which adaptive alignment should matter most. SDFA-Net records the best Dice and HD95 results on all three datasets and retains strong performance on irregular and small nodules.

\section{Related Work}

\subsection{Context-Aware Segmentation Architectures}

Convolutional Neural Networks (CNNs), particularly U-Net and its variants, are the foundation of medical image segmentation due to their ability to extract local texture features \cite{Ronneberger2015_e91a}. The symmetric encoder-decoder structure with skip connections preserves spatial details, which helps delineate the often ill-defined boundaries of thyroid nodules \cite{Zhou2018_4de5}. However, the locality of convolutional operations limits the effective receptive field, making it difficult to capture long-range anatomical dependencies and global semantic context \cite{Guo2019_0735}. Although methods like Attention U-Net attempt to expand the receptive field, they often rely on local attention windows, which may not fully capture irregular nodule morphologies \cite{Oktay2018_3460}. Consequently, CNN-based methods can struggle to distinguish nodules from surrounding tissues with similar echogenicity when global shape cues are missing \cite{Isensee2020_55c7}.

To address the constraints of fixed receptive fields, Vision Transformers (ViTs) and their hybrid variants have been rapidly adopted to use global self-attention mechanisms for medical image segmentation \cite{Dosovitskiy2020_b18a}. By modeling the relationships between all image patches, architectures such as TransUNet and Swin-Unet can effectively capture long-range anatomical dependencies, enabling a more comprehensive understanding of the global shape and structural context of lesions \cite{Chen2021_7e30,Cao2021_9b92}. This capability is theoretically advantageous for thyroid ultrasound segmentation, where the nodule's global shape often provides the most reliable cue against complex backgrounds. Furthermore, channel-wise cross-attention and multi-resolution feature synchronization have been used to bridge semantic gaps between encoder and decoder layers and strengthen multi-scale fusion \cite{Wang2021_7f7f,Chen2025S2DiNet,gao2024mba,gao2025cia}. These Transformer-based approaches demonstrate superior performance in scenarios requiring global semantic reasoning, moving beyond the inductive bias of locality that characterizes standard CNNs \cite{Hatamizadeh2021_97a5}.

Despite their global modeling capabilities, Transformer-based architectures have limitations in ultrasound imaging. The global self-attention mechanism treats all spatial tokens equally, potentially aggregating information from noisy background regions such as speckle artifacts, shadowing, or surrounding glandular tissues \cite{Li2024_33df}. This mixing of context can be detrimental in ultrasound images, where nodule-background contrast is low and acoustic noise is common, leading to boundary blurring or false positives \cite{Cai2023_6efa}. Moreover, the computational cost of quadratic self-attention often requires downsampling or token pruning, risking the loss of local details essential for boundary delineation \cite{Liu2021_09b7}. Consequently, current methods face a trade-off: while Transformers excel at global semantic understanding, they often lack the robustness to local noise found in CNNs. This highlights the need for architectures that selectively filter global context before fusing it with local geometric features \cite{Dai2021_21c0}.

\subsection{Semantic Filtering and Deformable Feature Fusion}

Integrating multi-scale features in medical segmentation requires navigating both semantic ambiguity and the need for precise spatial localization. Methods like attention gates achieve spatial filtering through gating coefficients, which selectively emphasize relevant feature regions based on local appearance compatibility \cite{Vanaja2025_3096}. While prototype learning focuses on deriving class-specific representations from feature statistics, thereby utilizing global distribution cues to distinguish categories \cite{Wang2019_b32a,Lv2026CPL}, attention mechanisms operate primarily by modulating the spatial importance of feature maps. This distinction highlights how attention gates address spatial compatibility by suppressing background noise through local weighting schemes, whereas prototype learning offers a robust statistical prior that captures class-specific characteristics from the global context \cite{Karri2022_91a0}.

Beyond semantic ambiguity, preserving boundary information across encoder and decoder features remains essential for accurate delineation \cite{Huang2026BFA,Pan2026MSBPNet,gao2025composite,gao2026camyla}. Standard fusion operations typically rely on fixed-grid convolution, which imposes rigid geometric constraints that fail to adapt to the irregular and non-rigid shapes characteristic of thyroid nodules \cite{Ronneberger2015_e91a}. In contrast, deformable convolutional networks (DCN) introduce learnable offsets that augment the sampling grid, allowing the receptive field to adjust adaptively to the object's geometry \cite{Dai2017_a4a3}. Content-aware networks use this adaptive nature by predicting offsets specific to the input content, effectively correcting geometric discrepancies that standard skip connections cannot resolve \cite{Dai2017_a4a3}. By replacing fixed sampling locations with deformable structures, these methods ensure that global semantic context is geometrically registered with local boundary details, leading to more precise feature integration \cite{Jaderberg2015_9523}.

\section{Method}

\subsection{Overview of SDFA-Net Architecture}

\begin{figure}[t]
\centering
\includegraphics[width=\columnwidth]{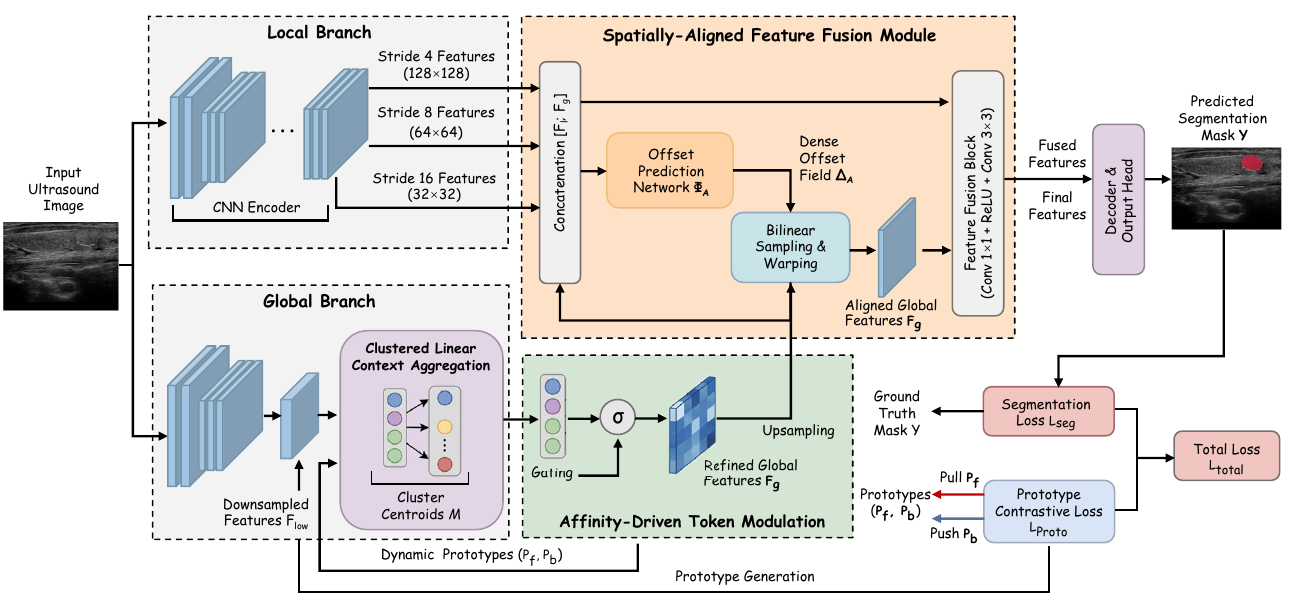}
\caption{Overall architecture of SDFA-Net. The model employs a dual-path design: a local path (blue) extracts high-resolution features, while a global context path (orange) captures long-range dependencies. The core 'filter-then-align' strategy consists of (1) Clustered Linear Context Aggregation and Prototype Modulation to filter semantically irrelevant features, and (2) Spatially-Aligned Feature Fusion to deformably align global context with local boundaries before decoding.}
\label{fig:method_fig1}
\end{figure}

SDFA-Net addresses thyroid nodule segmentation challenges by balancing local boundary delineation with global semantic understanding. As shown in Figure~\ref{fig:method_fig1}, our framework uses a dual-path architecture with a high-resolution local CNN branch and a context-aware global branch. The local branch, using a ResNet-34 backbone \cite{He2015_d11d}, preserves spatial details critical for tracing irregular nodule boundaries. Meanwhile, the global branch processes downsampled features to capture long-range anatomical dependencies and the overall thyroid context.

Our core contribution is the Semantic-Aware Deformable Feature Alignment pipeline, which uses a "filter-then-align" strategy. Instead of directly fusing raw multi-scale features—which often causes spatial misalignment and background noise interference—our approach first refines the global context using dynamic class prototypes. It then aligns the refined semantic priors with local features using learned deformable offsets. This two-stage process ensures that the final fusion integrates only semantically relevant information that is spatially registered to the local geometry.

In the first stage, the global branch transforms deep convolutional features into a sequence of tokens and applies a Clustered Linear Context Aggregation module to efficiently capture long-range dependencies with linear computational complexity. By grouping semantically similar tokens into learnable cluster centroids, this module approximates the effect of full self-attention while avoiding its quadratic cost. The resulting globally enriched tokens encode anatomical context spanning the entire thyroid region. However, these tokens may still contain interference from surrounding tissues and acoustic artifacts prevalent in ultrasound imaging. To address this, an Affinity-Driven Token Modulation mechanism uses dynamically computed class prototypes to filter out background noise. By calculating the semantic affinity of each token to the foreground nodule prototype, the module selectively suppresses irrelevant features while preserving those characteristic of the target structure.

Following semantic filtering, the refined global features must be integrated with the high-resolution local features extracted by the CNN backbone. Direct concatenation or addition assumes rigid spatial correspondence, which fails for thyroid nodules exhibiting irregular and deformable shapes. To overcome this limitation, the Spatially-Aligned Feature Fusion module employs a cascaded alignment strategy. It utilizes a hierarchical offset prediction network to estimate dense deformation fields at multiple scales. These content-aware offsets guide a differentiable bilinear sampling operation that iteratively warps the semantic context to align with the precise geometry encoded in multi-level local features. This deformable alignment ensures that semantic priors from the global path are geometrically consistent with boundary details from the local path at each resolution level.

The network is trained end-to-end using a composite loss function that combines pixel-level segmentation supervision with a prototype-based contrastive regularization. The segmentation loss employs both Binary Cross-Entropy and Dice loss to handle the severe class imbalance common in medical image segmentation. The prototype loss enforces semantic discriminability by pulling foreground tokens toward the foreground prototype while pushing background tokens away from the background prototype in the feature space. This dual supervision mechanism ensures that the network not only produces accurate segmentation masks but also learns a robust and semantically meaningful representation that supports effective noise filtering. The detailed formulation and implementation of each component are described in the following subsections.

\subsection{Dual-Path Feature Extraction and Context Aggregation}

Accurate segmentation of thyroid nodules demands a balance between high-resolution boundary delineation and comprehensive global context understanding. To achieve this, we employ a multi-scale dual-path extraction strategy based on a ResNet-34 backbone \cite{He2015_d11d}. The local path utilizes a feature pyramid extracted from multiple stages of the encoder with strides of 32, 16, 8, and 4, ensuring the preservation of fine-grained spatial details essential for tracing irregular nodule margins. Conversely, the global context path processes the deepest feature map from the final stage with a stride of 32. This lower-resolution input is subjected to average pooling, which significantly reduces the spatial dimensionality and computational load. By decoupling the paths, we maintain the geometric fidelity of multi-scale local features $\{F_{l1}, F_{l2}, F_{l3}\}$ while enabling the global branch to efficiently capture long-range anatomical dependencies from the downsampled representation $F_{low} \in \mathbb{R}^{C \times H_{32} \times W_{32}}$.

Although the global branch provides semantic context, standard self-attention mechanisms have $O(N^2)$ computational complexity \cite{Vaswani2017_e776}. To address this, we introduce the Clustered Linear Context Aggregation (CLCA) module, which reduces complexity to linear $O(NKC)$, where $K$ is the number of learnable cluster centroids and $C$ is the channel dimension. By grouping semantically similar tokens into clusters, we compress global information efficiently. The module calculates affinities only between tokens and learnable cluster centroids $M \in \mathbb{R}^{K \times C}$, modeling anatomical structures like nodule shape and position without the overhead of full self-attention.

The mathematical formulation of the CLCA module begins by computing the affinity matrix $A \in \mathbb{R}^{N \times K}$ between the input tokens $X \in \mathbb{R}^{N \times C}$ and the cluster centroids. We employ a differentiable Sinkhorn-Knopp normalization process to generate this affinity matrix. The doubly stochastic property enforced by the Sinkhorn algorithm ensures balanced assignments, preventing the collapse of clusters and significantly aiding the stability and convergence of the training process \cite{Cuturi2013_1f45}. Formally, the initial affinity score $\tilde{A}_{ij}$ is derived from the scaled dot-product similarity:
\begin{equation}
 \tilde{A}_{ij} = \frac{\exp(X_i M_j^\top / \tau)}{\sum_{k=1}^{K} \exp(X_i M_k^\top / \tau)},
\end{equation}
where $\tau$ is a learnable temperature parameter. The final affinity matrix is obtained as $A = \text{Sinkhorn}(\tilde{A})$. Using this matrix, the aggregated feature for the $j$-th cluster centroid $\hat{M}_j$ is computed as the weighted sum of all input tokens assigned to it:
\begin{equation}
 \hat{M}_j = \sum_{i=1}^{N} A_{ij} X_i.
\end{equation}
To reconstruct the global context for each individual token, the aggregated cluster features are broadcast back to the token space:
\begin{equation}
 G_i = \sum_{j=1}^{K} A_{ij} \hat{M}_j.
\end{equation}
Finally, the enhanced global tokens are produced via a residual connection:
\begin{equation}
 X_{global} = X + G.
\end{equation}

Following the context aggregation, the extracted local features $F_l$ and the enhanced global tokens $X_{global}$ must be integrated to support the final segmentation prediction. To prepare for the subsequent alignment and fusion steps, the global tokens $X_{global}$ are reshaped back into a 2D feature map $F_g \in \mathbb{R}^{C \times H_{32} \times W_{32}}$ and upsampled via bilinear interpolation to match the spatial resolution of the local features ($H_{16} \times W_{16}$). The fusion mechanism is initialized by concatenating the local and global feature maps along the channel dimension:
\begin{equation}
 F_{concat} = [F_l; F_{up}],
\end{equation}
where $F_{up}$ denotes the upsampled global features. This concatenated representation serves as the input for the deformable alignment module, which learns to spatially warp the global context to align with local boundaries before generating the final output.

\begin{figure}[t]
\centering
\includegraphics[width=0.7\columnwidth]{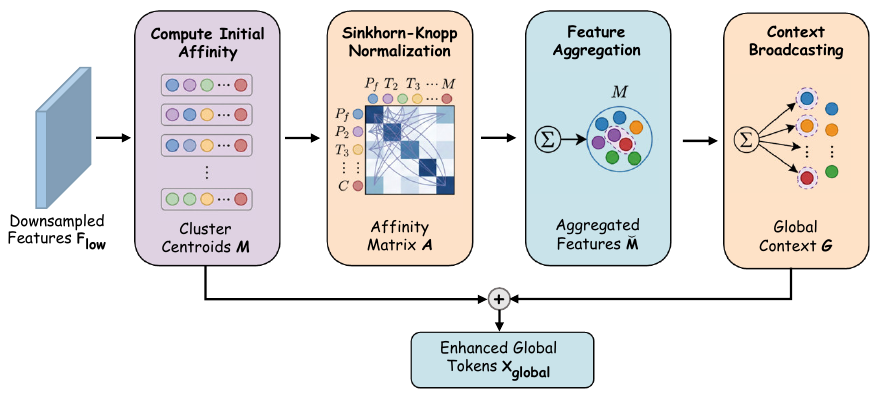}
\caption{Clustered Linear Context Aggregation (CLCA) module. The module groups semantically similar tokens into learnable cluster centroids via Sinkhorn normalization, capturing global anatomical context with linear computational complexity $O(NKC)$. The doubly stochastic affinity matrix ensures balanced cluster assignments, and aggregated features are broadcast back to reconstruct global context for each token.}
\label{fig:method_fig2}
\end{figure}

\subsection{Affinity-Driven Token Modulation}

\begin{figure}[t]
\centering
\includegraphics[width=0.7\columnwidth]{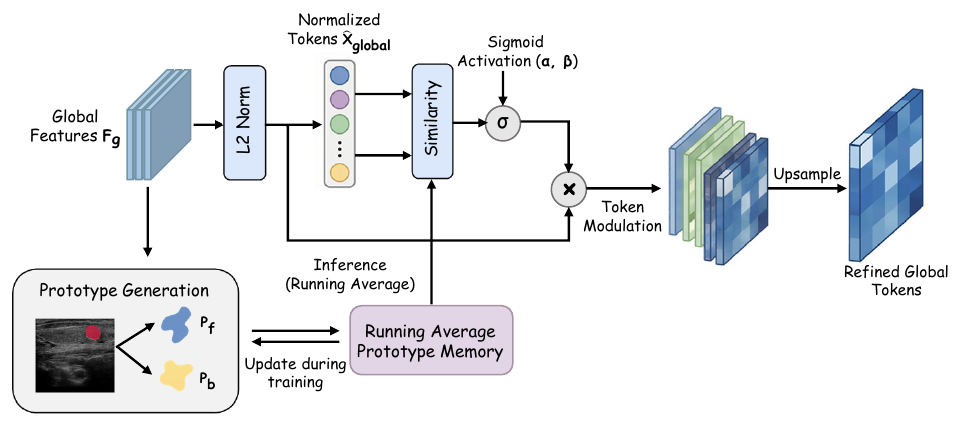}
\caption{Affinity-Driven Token Modulation mechanism. Dynamic class prototypes ($P_f$ for foreground, $P_b$ for background) are computed from L2-normalized global tokens using ground truth masks during training. Semantic affinity scores between each token and the foreground prototype are converted into gating coefficients via a learnable sigmoid function. These coefficients selectively suppress background tokens while preserving foreground features, producing semantically purified global context.}
\label{fig:method_fig3}
\end{figure}

Although clustered context aggregation captures long-range dependencies, global tokens may still encode irrelevant information from surrounding tissues like muscles or the trachea. To ensure only semantically meaningful features contribute to segmentation, we introduce an Affinity-Driven Token Modulation mechanism. This module filters background noise while amplifying nodule-related features. We use class-specific prototypes as anchors to compute the semantic affinity of each token. As shown in Figure~\ref{fig:method_fig3}, this process involves dynamically generating prototypes, calculating affinity scores, and applying a learned gating function to refine the global context.

To achieve robust semantic guidance, we compute class prototypes dynamically for each input image during training, using the ground truth mask $Y_{gt}$. Unlike static learnable prototypes \cite{Snell2017_9bca}, dynamic prototypes adapt to the specific anatomy and appearance of each scan, resolving potential circular dependencies in training. Let $X_{global} \in \mathbb{R}^{N \times C}$ denote the set of global tokens, where $N$ is the number of tokens and $C$ is the feature dimensionality. We first apply L2 normalization to the token vectors to obtain $\hat{X}_{global}$, ensuring that the affinity calculation focuses on semantic direction rather than magnitude. The foreground prototype $P_f$ and background prototype $P_b$ are then computed via masked average pooling over their respective spatial regions. Mathematically, the class-specific prototypes are defined as:

\begin{equation}
 P_c = \frac{1}{|\Omega_c|} \sum_{i \in \Omega_c} \hat{X}_{global,i}, \quad c \in \{f, b\}
\end{equation}
where $\Omega_f = \{i \mid Y_{gt,i} = 1\}$ represents the set of spatial indices belonging to the foreground nodule region, and $\Omega_b = \{i \mid Y_{gt,i} = 0\}$ denotes the background region.

Once the prototypes are established, we compute the semantic affinity of each global token to the foreground prototype and directly transform it into a gating coefficient. By using the L2-normalized vectors, the affinity score is computed using the dot product, which is equivalent to the cosine similarity measure. We then pass this score through a sigmoid activation function $\sigma$, parameterized by learnable scalars $\alpha$ and $\beta$, to obtain the gating coefficient $g_i$ for the $i$-th token:

\begin{equation}
 g_i = \sigma\left(\alpha \cdot (\hat{X}_{global,i} \cdot P_f^{\top}) + \beta\right)
\end{equation}
Here, the dot product $\hat{X}_{global,i} \cdot P_f^{\top}$ quantifies the degree of semantic relatedness between the token and the nodule region. The scaling factor $\alpha$ adjusts the sensitivity of the gate, effectively sharpening or softening the decision boundary between foreground and background, while the bias term $\beta$ shifts the threshold to allow the network to learn a prior preference for foreground or background activation. The final modulated token $X_{refined,i}$ is obtained by multiplying the original token $X_{global,i}$ by the gating coefficient:

\begin{equation}
 X_{refined,i} = g_i \cdot X_{global,i}
\end{equation}
This operation suppresses tokens with low affinity (likely background) and preserves those with high affinity (likely nodule), resulting in a purified global context representation.

During inference, the ground truth mask $Y_{gt}$ is unavailable, precluding the online calculation of dynamic prototypes. To address this, we maintain a running average of the foreground and background prototypes observed during the training phase. We update the prototype memory using an exponential moving average with a momentum coefficient $\mu$. This allows the model to generalize the learned semantic centroids to unseen data without requiring test-time annotations. The use of L2 normalization prior to affinity computation is essential, as it decouples the gating mechanism from variations in feature magnitude caused by different image contrasts or acquisition protocols, ensuring that the modulation relies solely on the semantic direction of the tokens. Consequently, the refined global features fed to the subsequent fusion module are highly discriminative and free from distracting background artifacts.

\subsection{Spatially-Aligned Feature Fusion}

Multi-scale feature integration is often hindered by spatial discrepancies between high-resolution local maps and coarse global context. Global features possess semantic priors but lack the geometric detail to delineate irregular boundaries. To bridge this gap, we propose a Spatially-Aligned Feature Fusion module that operates in a coarse-to-fine cascade. This module iteratively aligns and fuses high-level semantic features with multi-scale local features (strides 16, 8, and 4), inspired by feature pyramid networks and optical flow estimation \cite{Jaderberg2015_9523}.

\begin{figure}[t]
\centering
\includegraphics[width=0.7\columnwidth]{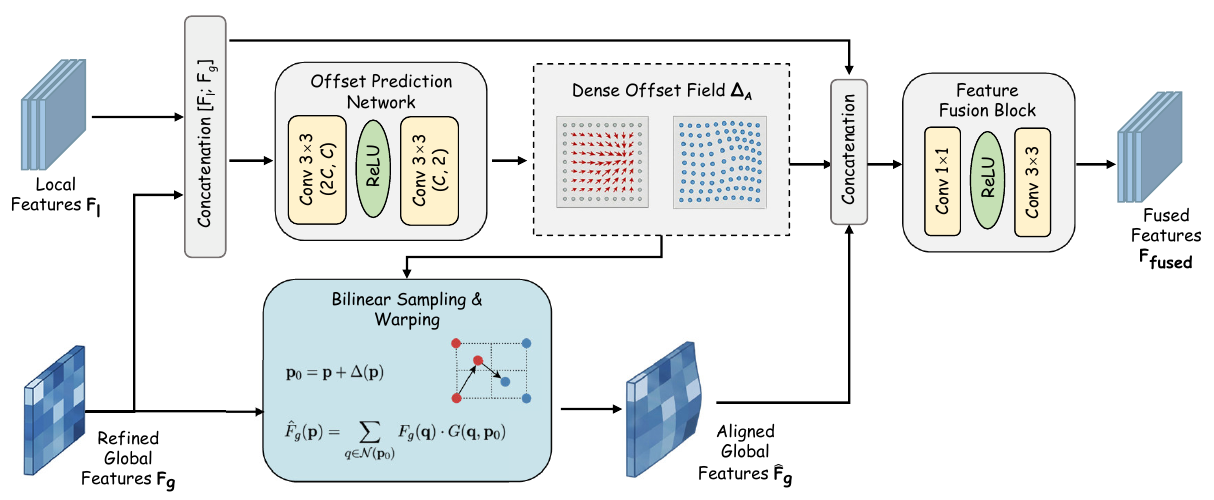}
\caption{Spatially-Aligned Feature Fusion module. The figure illustrates the alignment mechanism at a representative stage. An offset prediction network $\phi$ takes the concatenation of local features $F_l$ and upsampled semantic features to predict a dense 2D offset field $\Delta$. This field guides differentiable bilinear sampling to deformably warp the semantic features, producing aligned features that are geometrically registered to the local boundary details. This process is applied hierarchically across strides 16, 8, and 4.}
\label{fig:method_fig4}
\end{figure}

We treat feature fusion as a hierarchical content-aware warping task. By predicting dense 2D offset fields at each resolution stage, we allow the global semantic context to deform adaptively, aligning its structures with the fine-grained geometry of local features. This is particularly important for thyroid nodules with spiculated or lobulated margins, where rigid transformations fail. Figure~\ref{fig:method_fig4} illustrates the alignment mechanism at each stage.

Let $\{F_{l1}, F_{l2}, F_{l3}\}$ denote the multi-scale local features extracted at strides 4, 8, and 16, respectively. The alignment process begins with the refined global feature map $F_g$ (stride 32) and proceeds in a top-down manner. We define $F_{fused}^{(4)} = F_g$ as the initial semantic prior. For each stage $k \in \{3, 2, 1\}$ corresponding to strides $\{16, 8, 4\}$, the fusion process involves three steps: upsampling, offset prediction, and deformable warping.

First, the fused feature from the previous stage $F_{fused}^{(k+1)}$ is upsampled by a factor of 2 via bilinear interpolation to match the spatial resolution of the current local feature $F_{lk}$. An offset prediction network $\phi_k$ then estimates a dense displacement field $\Delta_k$ based on the concatenation of the local feature and the upsampled semantic prior:

\begin{equation}
\Delta_k = \phi_k([F_{lk}; \text{Up}(F_{fused}^{(k+1)})]).
\end{equation}

The network $\phi_k$ consists of two convolutional layers, predicting 2-channel offsets $(\Delta x, \Delta y)$ for each pixel. Using this offset field, we employ a differentiable bilinear sampling mechanism to warp the semantic prior, generating the aligned feature $\hat{F}_{semantic}^{(k)}$:

\begin{equation}
\hat{F}_{semantic}^{(k)}(p) = \text{Sampler}(\text{Up}(F_{fused}^{(k+1)}), p + \Delta_k(p)).
\end{equation}

Finally, the aligned semantic feature is fused with the local feature to produce the integrated representation for the current stage:

\begin{equation}
F_{fused}^{(k)} = \text{Conv}_{3\times3} (\text{ReLU}(\text{Conv}_{1\times1}([F_{lk}; \hat{F}_{semantic}^{(k)}]))).
\end{equation}

This cascaded process ensures that semantic priors are accurately registered to local details at each resolution level. The final output $F_{fused}^{(1)}$ at stride 4 serves as the input to the segmentation head, enabling precise boundary delineation even for small and irregular nodules.

\subsection{Loss Function and Training Objective}

To train the proposed SDFA-Net end-to-end, we formulate a composite objective function that addresses both pixel-level accuracy and the semantic discriminability of the intermediate features. Thyroid ultrasound segmentation is characterized by a severe class imbalance, where the nodule region often occupies a significantly smaller area than the background tissue. To ensure robust convergence and prevent the model from being biased toward the background class, we employ a hybrid segmentation loss $\mathcal{L}_{seg}$ that combines the complementary benefits of Binary Cross-Entropy (BCE) and the Dice similarity coefficient \cite{Sudre2017_393b}. The BCE loss provides stable gradients by evaluating pixel-wise probabilities independently, while the Dice loss directly optimizes the overlap metric, making the model highly sensitive to the shape and boundary of the nodule.

Formally, let $\hat{Y} \in \mathbb{R}^{H \times W}$ denote the predicted probability map generated by the decoder, and let $Y_{gt} \in \{0, 1\}^{H \times W}$ be the corresponding ground truth binary mask. The segmentation loss is defined as:

\begin{equation}
 \mathcal{L}_{seg} = \mathcal{L}_{bce}(\hat{Y}, Y_{gt}) + \mathcal{L}_{dice}(\hat{Y}, Y_{gt}),
\end{equation}

where the BCE term is computed as:

\begin{equation}
 \mathcal{L}_{bce} = - \frac{1}{HW} \sum_{i=1}^{HW} \left[ Y_{gt,i} \log(\hat{Y}_i) + (1 - Y_{gt,i}) \log(1 - \hat{Y}_i) \right],
\end{equation}

and the Dice loss is given by:

\begin{equation}
 \mathcal{L}_{dice} = 1 - \frac{2 \sum_{i=1}^{HW} \hat{Y}_i Y_{gt,i} + \epsilon}{\sum_{i=1}^{HW} \hat{Y}_i + \sum_{i=1}^{HW} Y_{gt,i} + \epsilon}.
\end{equation}

Here, $\epsilon$ is a smoothing constant introduced to prevent numerical instability.

While the segmentation loss supervises the final output, it does not explicitly enforce the semantic separation required for the effective operation of the Affinity-Driven Token Modulation module. To ensure that the global tokens used for prototype generation are semantically discriminative, we introduce a specialized prototype contrastive loss $\mathcal{L}_{proto}$ \cite{Li2020_6dce}. This loss acts as a regularizer that pulls foreground tokens towards the foreground prototype while pushing background tokens away from the background prototype in the feature space, thereby facilitating more accurate gating during the filtering process.

Given the set of normalized global tokens $\hat{X}_{global} \in \mathbb{R}^{N \times C}$ and the dynamically computed prototypes $P_f$ and $P_b$, the prototype contrastive loss is formulated as:

\begin{equation}
 \mathcal{L}_{proto} = \frac{1}{N} \sum_{i=1}^{N} \left[ Y_{gt,i} \cdot d(\hat{X}_{global,i}, P_f) + (1 - Y_{gt,i}) \cdot \max(0, m - d(\hat{X}_{global,i}, P_b)) \right],
\end{equation}

where $d(\cdot, \cdot)$ represents the cosine distance metric, and $m$ is a hyperparameter margin. For foreground tokens (where $Y_{gt,i}=1$), the loss minimizes the distance to the foreground prototype $P_f$. For background tokens (where $Y_{gt,i}=0$), the loss minimizes a hinge term that encourages the distance to the background prototype $P_b$ to be greater than the margin $m$, ensuring a distinct separation between semantic classes.

The final objective function for end-to-end training integrates the pixel-wise segmentation supervision with the semantic prototype regularization:

\begin{equation}
 \mathcal{L}_{total} = \mathcal{L}_{seg} + \lambda \mathcal{L}_{proto},
\end{equation}

where $\lambda$ is a weighting coefficient that balances the contribution of the prototype loss. We empirically set $\lambda$ to 0.1 to ensure that the contrastive learning provides sufficient semantic guidance without overwhelming the primary segmentation objective.

\section{Experiments}

\subsection{Datasets and Implementation Details}

We evaluated our approach on three public thyroid ultrasound benchmarks: TN3K \cite{gong2023thyroid}, DDTI \cite{pedraza2015open}, and TUCC\footnote{\url{https://stanfordaimi.azurewebsites.net/datasets/a72f2b02-7b53-4c5d-963c-d7253220bfd5}}. TN3K, our primary dataset, comprises approximately 3,500 static images from 208 patients, covering a wide variety of nodule sizes and echogenicity patterns. We partitioned TN3K into training (70\%), validation (10\%), and test (20\%) sets. The DDTI dataset, derived from 61 patient video sequences, was similarly partitioned (70/10/20) to maintain consistency. The TUCC dataset contains cine-clips acquired with different equipment and protocols, providing a complementary evaluation on data with distinct characteristics; it was also partitioned using the 70/10/20 split. All images were resized to $512 \times 512$ pixels and normalized. These datasets contain nodules with highly irregular boundaries, making them suitable benchmarks for boundary-sensitive segmentation.

Quantitative assessment was performed using standard metrics that capture both volumetric overlap and boundary fidelity. The primary metric is the Dice Similarity Coefficient (DSC), which measures the intersection between the predicted segmentation and the ground truth relative to their average size:
\begin{equation}
\text{DSC}(P, G) = \frac{2|P \cap G|}{|P| + |G|}.
\end{equation}
While DSC provides a global measure of segmentation accuracy, it can be insensitive to local boundary deviations. Therefore, we also utilized the 95th percentile Hausdorff Distance (HD95). HD95 calculates the maximum surface distance between the predicted and ground truth boundaries after removing the top 5\% of outliers to ensure robustness against noise. Lower HD95 values indicate better alignment with the actual nodule margin, which is particularly important for identifying spiculated or invasive boundaries associated with malignant tumors.

The proposed framework was implemented using the PyTorch library and trained on a single NVIDIA RTX 4090 GPU. We employed a ResNet-34, pre-trained on ImageNet \cite{deng2009imagenet}, as the backbone for the local feature extraction path to benefit from transfer learning from natural images, while the global context branch was initialized with random weights. The network was optimized end-to-end using the AdamW optimizer with an initial learning rate of $1 \times 10^{-4}$ and a weight decay of 0.05. We employed a polynomial decay schedule to adjust the learning rate dynamically over 120 training epochs. The model was trained using a composite loss function combining Binary Cross-Entropy (BCE) and Dice loss to handle class imbalance, supplemented by the prototype contrastive loss to enforce semantic discrimination. To enhance the model's robustness to ultrasound imaging artifacts and orientation variations, we applied intensive data augmentation during training, including random rotations within the range of $[-90^{\circ}, 90^{\circ}]$, horizontal and vertical flipping, and random adjustments to brightness and contrast. Additionally, we utilized elastic deformations and Gaussian noise injection to simulate the speckle noise present in ultrasound imaging. The number of cluster centroids $K$ in the context aggregation module was set to 8, and the weight for the prototype loss $\lambda$ was empirically set to 0.1. The final batch size was fixed at 16 to balance GPU memory consumption with training stability.

\subsection{Ablation Studies}

We conduct comprehensive ablation studies on the primary dataset (TN3K) to validate the contribution of each proposed module, analyze design alternatives, and assess the sensitivity of key hyperparameters.

\subsubsection{Component Ablations}

Table~\ref{tab:ablation-component} presents a progressive evaluation of the core components in SDFA-Net. Starting from a U-Net baseline, we incrementally add the Clustered Linear Context Aggregation (Ctx), Affinity-Driven Token Modulation (Mod), and Spatially-Aligned Feature Fusion (Align). The baseline model achieves a Dice score of $80.79\pm1.85\%$ with an HD95 of $21.51$ pixels. Introducing the context aggregation module significantly boosts performance by $+2.67\%$ in Dice and reduces HD95 by $-4.68$ pixels, demonstrating the necessity of capturing long-range anatomical dependencies. Further incorporating the modulation mechanism yields another $+2.27\%$ improvement in Dice and lowers HD95 by $-2.61$ pixels, effectively filtering out background noise. Finally, adding the deformable alignment module results in the full SDFA-Net, which achieves the best performance with $87.42\pm0.96\%$ Dice and the lowest HD95 of $12.37$ pixels. This confirms that the synergistic combination of semantic filtering and geometric alignment is crucial for precise boundary delineation.

\begin{table}[htbp]
\centering
\caption{Component ablation results on the TN3K validation set.}
\label{tab:ablation-component}
\begin{tabular}{lcc}
\toprule
Configuration & Dice (\%) $\uparrow$ & HD95 (px) $\downarrow$ \\
\midrule
Base (U-Net) & 80.79 $\pm$ 1.85 & 21.51 $\pm$ 3.42 \\
+ Ctx & 83.46 $\pm$ 1.41 & 16.83 $\pm$ 2.71 \\
+ Ctx + Mod & 85.73 $\pm$ 1.19 & 14.22 $\pm$ 2.35 \\
\textbf{Full (SDFA-Net)} & \textbf{87.42 $\pm$ 0.96} & \textbf{12.37 $\pm$ 2.04} \\
\bottomrule
\end{tabular}
\end{table}

\subsubsection{Global Context Aggregation Method}

To validate the efficiency and design of the Clustered Linear Context Aggregation module, we compare it against standard Global Average Pooling (GAP) and full self-attention in Table~\ref{tab:ablation-context}. We also analyze the impact of varying the number of clusters $K$. GAP performs the worst ($82.08\%$ Dice) due to the complete loss of spatial structure. While standard self-attention achieves a Dice score comparable to our method ($83.51\%$ vs. $83.46\%$), it incurs a significantly higher computational cost of $48.73$ GFLOPs, which is approximately $39\%$ more than our clustered approach. Our clustered method with $K=8$ achieves the optimal trade-off, balancing performance ($83.46\%$ Dice, $16.83$ px HD95) with efficiency ($35.11$ GFLOPs). Performance saturation is observed when increasing $K$ to $16$, indicating that a limited number of clusters is sufficient to model the global context in thyroid ultrasound images effectively.

\begin{table}[htbp]
\centering
\caption{Comparison of global context aggregation strategies. Params denotes parameters in millions.}
\label{tab:ablation-context}
\resizebox{\linewidth}{!}{
\begin{tabular}{lcccc}
\toprule
Variant & Dice (\%) $\uparrow$ & HD95 (px) $\downarrow$ & Params (M) & GFLOPs \\
\midrule
Global Average Pooling (GAP) & 82.08 $\pm$ 1.52 & 18.94 $\pm$ 3.02 & 24.87 & 32.15 \\
Standard Self-Attention & 83.51 $\pm$ 1.39 & 16.75 $\pm$ 2.68 & 25.12 & 48.73 \\
Clustered ($K=4$) & 83.12 $\pm$ 1.44 & 17.41 $\pm$ 2.79 & 24.93 & 34.82 \\
\textbf{Clustered ($K=8$)} & \textbf{83.46 $\pm$ 1.41} & \textbf{16.83 $\pm$ 2.71} & \textbf{24.96} & \textbf{35.11} \\
Clustered ($K=16$) & 83.39 $\pm$ 1.42 & 16.91 $\pm$ 2.73 & 25.02 & 35.69 \\
\bottomrule
\end{tabular}}
\end{table}

\subsubsection{Prototype Modulation Strategy}

This experiment evaluates the necessity of the proposed dynamic prototype generation and gating mechanism. As shown in Table~\ref{tab:ablation-modulation}, we compare the proposed dynamic prototypes against a configuration with no modulation and one using static, learnable prototypes. Removing modulation ($g_i=1$) leads to a significant drop of $-2.27\%$ Dice on TN3K, confirming that unfiltered global context introduces noise that degrades segmentation accuracy. Static prototypes improve performance over no modulation but lag behind our dynamic approach by $-0.81\%$ Dice. Crucially, the advantage of dynamic prototypes is consistent across different datasets, including TUCC where they achieve a $+2.04\%$ Dice improvement over static prototypes. This demonstrates that dynamic prototypes adapt to the specific semantics of each input, providing more robust and generalizable feature filtering across datasets with varying imaging characteristics.

\begin{table}[htbp]
\centering
\caption{Ablation study on prototype modulation strategies, evaluated on TN3K validation set and TUCC test set.}
\label{tab:ablation-modulation}
\begin{tabular}{lccc}
\toprule
Variant & Dice (\%) $\uparrow$ & HD95 (px) $\downarrow$ & TUCC (Test Dice) \\
\midrule
No Modulation ($g_i = 1$) & 83.46 $\pm$ 1.41 & 16.83 $\pm$ 2.71 & 66.54 $\pm$ 2.47 \\
Static Learnable Prototypes & 84.92 $\pm$ 1.27 & 15.04 $\pm$ 2.48 & 68.19 $\pm$ 2.32 \\
\textbf{Dynamic Prototypes (Ours)} & \textbf{85.73 $\pm$ 1.19} & \textbf{14.22 $\pm$ 2.35} & \textbf{70.23 $\pm$ 2.45} \\
\bottomrule
\end{tabular}
\end{table}

\subsubsection{Feature Fusion Strategy}

Table~\ref{tab:ablation-fusion} isolates the contribution of the Spatially-Aligned Feature Fusion module by comparing our deformable alignment strategy against simple concatenation and a global rigid (affine) alignment. Direct concatenation results in the highest HD95 ($14.22$ pixels), as global features are not spatially registered to local boundary details, leading to misalignment artifacts. Rigid alignment improves both Dice ($+0.55\%$) and HD95 ($-0.57$ pixels) by correcting for global positional shifts. However, our deformable alignment provides the most substantial improvement, increasing Dice by an additional $+1.14\%$ and achieving the lowest HD95 of $12.37$ pixels. This confirms that dense, content-aware offsets are necessary to handle the non-rigid and highly irregular shapes characteristic of thyroid nodules.

\begin{table}[htbp]
\centering
\caption{Comparison of different feature fusion strategies.}
\label{tab:ablation-fusion}
\begin{tabular}{lcc}
\toprule
Fusion Strategy & Dice (\%) $\uparrow$ & HD95 (px) $\downarrow$ \\
\midrule
Direct Concatenation & 85.73 $\pm$ 1.19 & 14.22 $\pm$ 2.35 \\
Rigid (Affine) Alignment & 86.28 $\pm$ 1.08 & 13.65 $\pm$ 2.22 \\
\textbf{Deformable Alignment (Ours)} & \textbf{87.42 $\pm$ 0.96} & \textbf{12.37 $\pm$ 2.04} \\
\bottomrule
\end{tabular}
\end{table}

\subsubsection{Hyperparameter Sensitivity}

The prototype contrastive loss $\mathcal{L}_{proto}$ serves to supervise the semantic space by enforcing separation between foreground and background features. We analyze the sensitivity of the model to the weight $\lambda$ of this loss in the total objective $\mathcal{L}_{total} = \mathcal{L}_{seg} + \lambda \mathcal{L}_{proto}$. Table~\ref{tab:ablation-hyper} shows that the model is sensitive to this hyperparameter. Removing the prototype loss ($\lambda=0.00$) causes a noticeable performance drop of $-0.55\%$ in Dice and an increase of $+0.61$ pixels in HD95 compared to the optimal setting. Performance peaks at $\lambda=0.10$, yielding $87.42\%$ Dice and $12.37$ pixels HD95. Increasing $\lambda$ beyond $0.10$ leads to diminishing returns and slight degradation, likely because an overly strong contrastive loss begins to interfere with the primary segmentation objective. This confirms that a moderate amount of explicit semantic space supervision is beneficial for stable convergence and high accuracy.

\begin{table}[htbp]
\centering
\caption{Sensitivity analysis of the prototype loss weight $\lambda$.}
\label{tab:ablation-hyper}
\begin{tabular}{lcc}
\toprule
$\lambda$ & Dice (\%) $\uparrow$ & HD95 (px) $\downarrow$ \\
\midrule
0.00 (No $\mathcal{L}_{proto}$) & 86.87 $\pm$ 1.02 & 12.98 $\pm$ 2.18 \\
0.05 & 87.18 $\pm$ 0.99 & 12.59 $\pm$ 2.09 \\
\textbf{0.10} & \textbf{87.42 $\pm$ 0.96} & \textbf{12.37 $\pm$ 2.04} \\
0.20 & 87.31 $\pm$ 0.97 & 12.45 $\pm$ 2.06 \\
0.50 & 86.95 $\pm$ 1.01 & 12.83 $\pm$ 2.15 \\
\bottomrule
\end{tabular}
\end{table}

\subsection{State-of-the-Art Comparisons}

We compared SDFA-Net with state-of-the-art methods on the TN3K, TUCC, and DDTI datasets. To demonstrate generalizability, all datasets were evaluated using supervised training with a consistent 70/10/20 split. Performance was quantified using the Dice Similarity Coefficient (DSC) and the 95th percentile Hausdorff Distance (HD95).

\subsubsection{Performance on TN3K}
TN3K served as our primary benchmark. As shown in Table~\ref{tab:sota_tn3k}, SDFA-Net achieved a Dice score of 87.42\%, outperforming the strongest baseline, AVFM \cite{zhang2025adapting}, by 1.41\%. Crucially, our method reduced HD95 to 12.37 pixels, a significant improvement over AVFM's 15.42 pixels. This reduction in boundary error validates the Spatially-Aligned Feature Fusion module's ability to delineate nodule margins. Deformable alignment allows the model to correct geometric misalignments between global semantic context and local features, which is critical for irregular shapes.

\begin{table}[htbp]
\centering
\caption{State-of-the-art comparison on TN3K dataset.}
\label{tab:sota_tn3k}
\resizebox{\linewidth}{!}{
\begin{tabular}{llcc}
\toprule
Method & Venue & Dice (\%) $\uparrow$ & HD95 (px) $\downarrow$ \\
\midrule
U-Net \cite{Ronneberger2015_e91a} & MICCAI 2015 & 80.79 $\pm$ 1.85 & 21.51 $\pm$ 3.42 \\
Attention U-Net \cite{Oktay2018_3460} & MIDL 2018 & 81.25 $\pm$ 1.72 & 19.05 $\pm$ 2.91 \\
TransUNet \cite{Chen2021_7e30} & MICCAI 2021 & 80.98 $\pm$ 1.78 & 18.81 $\pm$ 3.11 \\
Swin-Unet \cite{Cao2021_9b92} & ECCVW 2022 & 77.18 $\pm$ 2.04 & 57.28 $\pm$ 8.43 \\
nnU-Net \cite{Isensee2020_55c7} & Nat. Methods 2021 & 85.13 $\pm$ 1.12 & 16.05 $\pm$ 2.87 \\
UCTransNet \cite{Wang2021_7f7f} & AAAI 2022 & 81.99 $\pm$ 1.54 & 43.36 $\pm$ 5.62 \\
BPAT-UNet \cite{bi2023bpat} & CMPB 2023 & 83.64 $\pm$ 1.38 & 14.06 $\pm$ 2.33 \\
HFA-UNet \cite{li2025hfa} & KBS 2025 & 84.58 $\pm$ 1.21 & 38.85 $\pm$ 4.97 \\
CC-SAM \cite{gowda2024cc} & ECCV 2024 & 85.20 $\pm$ 1.08 & 27.10 $\pm$ 3.82 \\
AVFM \cite{zhang2025adapting} & MICCAI 2025 & 86.01 $\pm$ 1.03 & 15.42 $\pm$ 2.51 \\
SDFA-Net (Ours) & - & \textbf{87.42 $\pm$ 0.96} & \textbf{12.37 $\pm$ 2.04} \\
\bottomrule
\end{tabular}}
\end{table}

\begin{figure}[t]
\centering
\includegraphics[width=\linewidth]{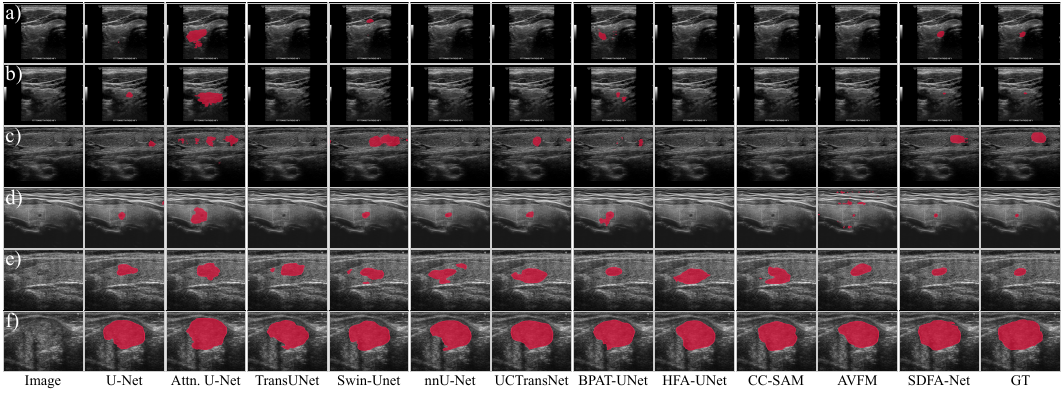}
\caption{Qualitative segmentation results of SDFA-Net on three datasets: TN3K (a, b), TUCC (c, d), and DDTI (e, f). The predicted segmentation masks are overlaid in red on the original ultrasound images. Our method demonstrates superior capability in delineating irregular boundaries and suppressing background noise across diverse scanning protocols.}
\label{fig:visual_results}
\end{figure}

Furthermore, analysis of nodule subtypes reveals that our model maintains robust performance across varying geometries. As illustrated in Figure~\ref{fig:visual_results}, we observe a distinct advantage in segmenting irregular nodules, which often pose the greatest challenge due to spiculated margins. By filtering semantically irrelevant features through prototype modulation, the network focuses on the nodule core, preventing background interference from distorting the boundary prediction.

\subsubsection{Performance on TUCC}
The TUCC dataset was partitioned using the same 70/10/20 protocol and trained independently to evaluate performance on data acquired with different ultrasound equipment and imaging protocols. The results in Table~\ref{tab:sota_tucc} demonstrate that our method maintains strong performance on this dataset despite the distinct imaging characteristics. SDFA-Net attains a Dice score of 71.83\%, surpassing the best cross-dataset baseline, AVFM \cite{zhang2025adapting}, by 2.96\%. This performance demonstrates the robustness of our semantic filtering mechanism across varying acquisition protocols. The Affinity-Driven Token Modulation module effectively suppresses protocol-specific artifacts and background textures, allowing the model to focus on generalizable nodule features rather than dataset-specific appearance patterns.

\begin{table}[htbp]
\centering
\caption{State-of-the-art comparison on TUCC dataset.}
\label{tab:sota_tucc}
\resizebox{\linewidth}{!}{
\begin{tabular}{llcc}
\toprule
Method & Venue & Dice (\%) $\uparrow$ & HD95 (px) $\downarrow$ \\
\midrule
U-Net \cite{Ronneberger2015_e91a} & MICCAI 2015 & 44.08 $\pm$ 3.72 & 186.51 $\pm$ 25.43 \\
Attention U-Net \cite{Oktay2018_3460} & MIDL 2018 & 46.33 $\pm$ 3.58 & 172.84 $\pm$ 23.67 \\
TransUNet \cite{Chen2021_7e30} & MICCAI 2021 & 45.96 $\pm$ 3.65 & 124.03 $\pm$ 19.54 \\
Swin-Unet \cite{Cao2021_9b92} & ECCVW 2022 & 55.02 $\pm$ 3.12 & 106.65 $\pm$ 16.38 \\
nnU-Net \cite{Isensee2020_55c7} & Nat. Methods 2021 & 64.61 $\pm$ 2.87 & 40.05 $\pm$ 6.02 \\
UCTransNet \cite{Wang2021_7f7f} & AAAI 2022 & 58.19 $\pm$ 2.98 & 87.98 $\pm$ 13.45 \\
BPAT-UNet \cite{bi2023bpat} & CMPB 2023 & 52.87 $\pm$ 3.28 & 89.23 $\pm$ 13.71 \\
HFA-UNet \cite{li2025hfa} & KBS 2025 & 63.73 $\pm$ 2.67 & 87.33 $\pm$ 12.94 \\
CC-SAM \cite{gowda2024cc} & ECCV 2024 & 65.18 $\pm$ 2.54 & 75.42 $\pm$ 11.28 \\
AVFM \cite{zhang2025adapting} & MICCAI 2025 & 68.87 $\pm$ 2.38 & 30.85 $\pm$ 5.63 \\
SDFA-Net (Ours) & - & \textbf{71.83 $\pm$ 2.21} & \textbf{26.14 $\pm$ 4.82} \\
\bottomrule
\end{tabular}}
\end{table}

\subsubsection{Performance on DDTI}
We validate SDFA-Net on the DDTI dataset using the same 70/10/20 training, validation, and test split. This dataset includes nodules with significant boundary ambiguity. As reported in Table~\ref{tab:sota_ddti}, our method achieves a Dice score of 86.73\%, outperforming the previous best method, HFA-UNet \cite{li2025hfa}, by 1.53\%. The HD95 is drastically reduced to 11.05 pixels, confirming that the Clustered Linear Context Aggregation module effectively uses global anatomical context to resolve local ambiguities. Notably, the improvement is most pronounced on nodules with low annotation quality, where the standard deviation of Dice is higher across baselines. Our prototype-guided approach stabilizes predictions in these uncertain regions by enforcing consistency with learned semantic priors.

\begin{table}[htbp]
\centering
\caption{State-of-the-art comparison on DDTI dataset.}
\label{tab:sota_ddti}
\resizebox{\linewidth}{!}{
\begin{tabular}{llcc}
\toprule
Method & Venue & Dice (\%) $\uparrow$ & HD95 (px) $\downarrow$ \\
\midrule
U-Net \cite{Ronneberger2015_e91a} & MICCAI 2015 & 71.82 $\pm$ 3.47 & 40.70 $\pm$ 7.35 \\
Attention U-Net \cite{Oktay2018_3460} & MIDL 2018 & 73.28 $\pm$ 3.32 & 38.42 $\pm$ 6.88 \\
TransUNet \cite{Chen2021_7e30} & MICCAI 2021 & 74.83 $\pm$ 3.18 & 31.98 $\pm$ 5.94 \\
Swin-Unet \cite{Cao2021_9b92} & ECCVW 2022 & 79.85 $\pm$ 2.49 & 24.80 $\pm$ 4.63 \\
nnU-Net \cite{Isensee2020_55c7} & Nat. Methods 2021 & 82.84 $\pm$ 1.87 & 16.23 $\pm$ 3.28 \\
UCTransNet \cite{Wang2021_7f7f} & AAAI 2022 & 80.75 $\pm$ 2.61 & 24.96 $\pm$ 4.71 \\
BPAT-UNet \cite{bi2023bpat} & CMPB 2023 & 78.95 $\pm$ 2.71 & 22.18 $\pm$ 4.25 \\
HFA-UNet \cite{li2025hfa} & KBS 2025 & 85.20 $\pm$ 1.63 & 23.90 $\pm$ 4.52 \\
CC-SAM \cite{gowda2024cc} & ECCV 2024 & 81.52 $\pm$ 2.14 & 26.68 $\pm$ 4.89 \\
AVFM \cite{zhang2025adapting} & MICCAI 2025 & 83.30 $\pm$ 1.92 & 13.23 $\pm$ 2.87 \\
SDFA-Net (Ours) & - & \textbf{86.73 $\pm$ 1.48} & \textbf{11.05 $\pm$ 2.41} \\
\bottomrule
\end{tabular}}
\end{table}

\begin{figure}[t]
\centering
\includegraphics[width=\linewidth]{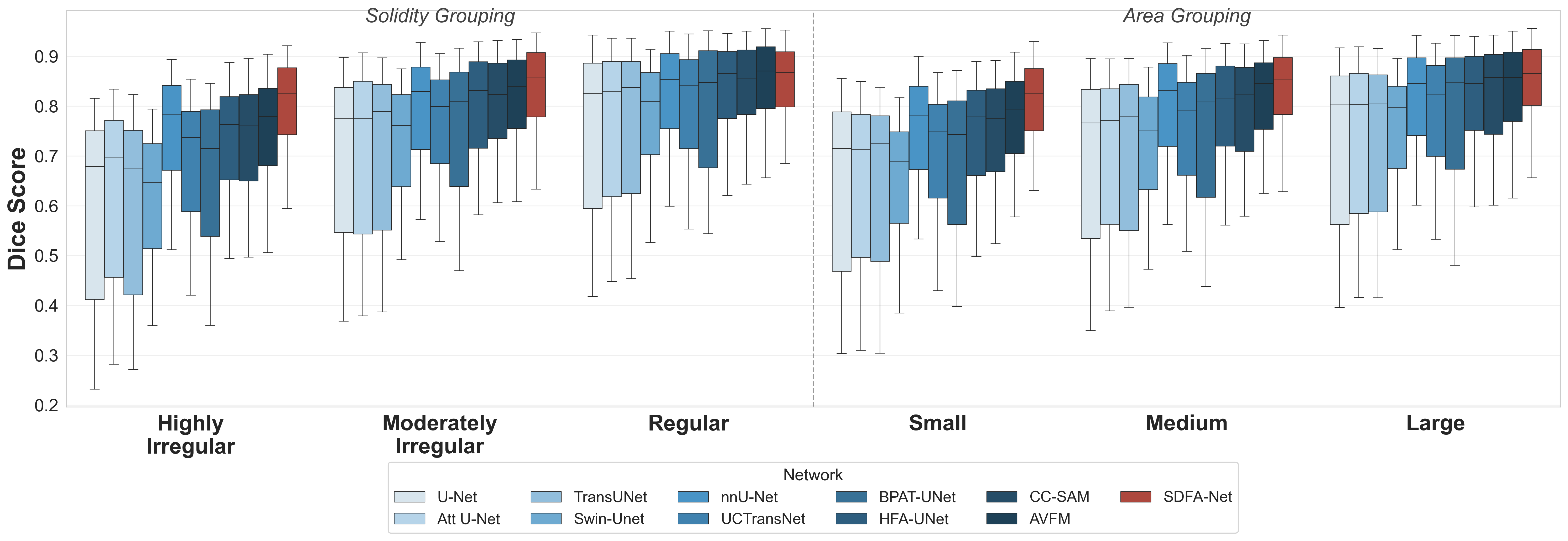}
\caption{Robustness analysis across nodule solidity and size groups on TN3K, TUCC, and DDTI datasets. SDFA-Net (red) consistently outperforms baselines, particularly in challenging groups (Highly Irregular, Small), demonstrating superior generalizability.}
\label{fig:robustness}
\end{figure}

\subsection{Robustness Analysis}

To further investigate the robustness of SDFA-Net against geometric variations, we conduct a fine-grained performance analysis based on nodule irregularity and size. We employ \textit{solidity} as a quantitative metric to measure shape irregularity, defined as the ratio of the nodule's area to the area of its convex hull:
\begin{equation}
\text{solidity} = \frac{\text{Area}}{\text{ConvexHullArea}}
\end{equation}
where a value close to 1 indicates a regular, convex shape, while lower values signify irregular, lobulated, or spiculated boundaries commonly associated with malignancy \cite{soyer2021impact,tessler2017acr}.

We stratify the test samples from all three datasets into three equal-sized subgroups (Highly Irregular, Moderately Irregular, Regular) based on the terciles of their solidity scores. Similarly, we group nodules by size (Small, Medium, Large) based on the terciles of their pixel area, ensuring balanced sample sizes for robust statistical analysis. Figure~\ref{fig:robustness} presents the Dice score distributions for each subgroup.

As visually summarized in Figure~\ref{fig:robustness}, SDFA-Net consistently outperforms all baseline methods across different solidity and size groups. In the ``Highly Irregular'' group, which represents the most challenging cases with complex boundaries, our method achieves a significant performance gain, validating the efficacy of the Spatially-Aligned Feature Fusion module in capturing fine-grained geometric details. Similarly, for ``Small'' nodules, SDFA-Net maintains high accuracy where other methods struggle due to limited spatial resolution. The consistent reduction in HD95 across all three datasets (Tables~\ref{tab:sota_tn3k}-\ref{tab:sota_ddti}) further underscores the clinical relevance of our approach, as minimizing boundary errors is critical for reliable diagnosis.

\section{Discussion}

Results across three datasets support a consistent account of SDFA-Net. Prototype filtering reduces semantic noise before fusion, and spatial alignment places the remaining context at locations that agree with local boundary evidence. The gain appears in both region overlap and boundary distance, with strong value in irregular and small nodules. This pattern connects the final accuracy to the intended roles of the two stages and supports semantic selection followed by geometric matching as the main design principle.

From a pattern recognition perspective, the contribution lies in the order imposed on multiscale reasoning. The model first estimates which global tokens belong to the target class, then estimates where that information should enter the local feature map. This order prevents noisy context from steering the offset field and gives the alignment stage a cleaner semantic target. The same principle can apply to dense recognition problems in which class evidence is stable at low resolution while object geometry varies at high resolution.

From a clinical perspective, the real-time inference capability of approximately 23 frames per second makes SDFA-Net a viable candidate for integration into computer-aided diagnosis systems. The robust handling of speckle noise and shadowing artifacts reduces the need for manual correction by radiologists, potentially streamlining the diagnostic workflow. Furthermore, the model's consistent strong performance across datasets acquired with different ultrasound equipment and imaging protocols suggests that the learned semantic representations are not overfitted to specific machine characteristics, increasing the utility of the framework across different clinical centers with varying imaging systems.

The current experiments cover fully supervised 2D ultrasound segmentation across three public datasets. This scope isolates the effect of semantic filtering and spatial alignment under consistent training conditions. Future work will study sparse annotation, volumetric ultrasound, and deployment specific optimization. These settings will test whether the same feature alignment principle remains effective with weaker supervision, richer spatial context, and tighter computation budgets.

\section{Conclusion}

SDFA-Net addresses a central difficulty in thyroid ultrasound segmentation. Reliable prediction requires global evidence that identifies the nodule and local evidence that preserves its irregular boundary. The proposed pipeline processes these signals in a clear order. Clustered linear context aggregation builds an efficient global representation. Dynamic prototypes remove context that does not match the nodule class. Cascaded deformable fusion then registers the filtered representation with local features across three resolutions. This sequence gives the decoder semantic context that agrees with the geometry of the target.

Results on TN3K, TUCC, and DDTI support this design. SDFA-Net achieves the highest Dice score and the lowest HD95 among the compared methods on all three datasets. Component studies trace these gains to the context, prototype, and alignment modules. Shape and size analysis shows a strong benefit for highly irregular and small nodules, where fixed spatial fusion is most likely to miss boundary structure. The model also runs at about 23 frames per second. These findings make semantic filtering followed by geometric alignment a useful design principle for boundary sensitive recognition.

The current study covers fully supervised 2D ultrasound segmentation. Future work will study sparse labels, 3D ultrasound, and deployment specific optimization. These directions will test the same principle under weaker supervision, volumetric context, and tighter computation budgets. The broader value of SDFA-Net lies in treating semantic relevance and spatial correspondence as two linked decisions. This view offers a direct route for dense prediction tasks in which global class evidence and local object geometry appear at different scales.

\section*{CRediT authorship contribution statement}
\textbf{Lei Zhang:} Conceptualization, Methodology, Software, Writing -- original draft.
\textbf{Ruhuan Zhou:} Software, Validation, Data curation, Writing -- original draft.
\textbf{Busheng Tong:} Methodology, Data curation.
\textbf{Dan Lv:} Conceptualization, Visualization.
\textbf{Bingyu Liang:} Visualization.
\textbf{Taibao Wang:} Writing -- original draft.
\textbf{Xinyu Shi:} Supervision.
\textbf{Yifan Gao:} Software, Visualization.
\textbf{Ming Zhang:} Writing -- review \& editing, Supervision.
\textbf{Qin Wang:} Writing -- review \& editing, Supervision, Funding acquisition.

\section*{Declaration of competing interest}
The authors declare that they have no known competing financial interests or personal relationships that could have appeared to influence the work reported in this paper.

\section*{Data availability}
All datasets analyzed in this study are publicly available: TN3K \cite{gong2023thyroid} (\href{https://github.com/haifangong/TRFE-Net-for-thyroid-nodule-segmentation}{dataset repository}), DDTI \cite{pedraza2015open} (\href{https://www.kaggle.com/datasets/dasmehdixtr/ddti-thyroid-ultrasound-images}{dataset page}), and TUCC (\href{https://stanfordaimi.azurewebsites.net/datasets/a72f2b02-7b53-4c5d-963c-d7253220bfd5}{dataset page}). No new dataset was generated in this study.

\section*{Declaration of generative AI and AI-assisted technologies in the manuscript preparation process}
During the preparation of this work, the authors used OpenAI Codex to assist with language refinement, document organization, and compliance checking. After using this tool, the authors reviewed and edited the content as needed and take full responsibility for the content of the publication.

\section*{Acknowledgments}
This work was supported by the General Programs of the Anhui Natural Science Foundation under Grant 2508085MH207; the Young Scientists Program of the National Natural Science Foundation of China under Grant 81700898; the Provincial Special Project for Clinical Medicine Research and Translation under Grant 202527c10020032; the Annual Applied Medical Research Project of Hefei Municipal Health Commission under Grant Hwk2023zc002; and the Member units and non-directly affiliated hospitals of the Health Alliance, Science and Technology Project of Bengbu Medical University under Grant 2024byzd376. The funders had no role in study design, data collection, data analysis or interpretation, manuscript preparation, or the decision to submit the work for publication.

\bibliographystyle{elsarticle-num}
\bibliography{references}

\end{document}